\documentclass[twoside]{article}

\usepackage{graphicx}

\usepackage{cmpj2e}

\title[Spin fluctuations and the "strange metal" behavior]%
{Spin fluctuations and the "strange metal" behavior of a weakly doped antiferromagnet%
\thanks{It is a great pleasure for us to contribute with this paper to a special issue of Condensed
Matter Physics on the occasion of the 70th anniversary of a
well-known Ukrainian theoretical physicist, a Corresponding Member
of The National Academy of Sciences of Ukraine, Professor
Ihor~V.~Stasyuk, who contributed in both fields of the electronic properties of strongly
correlated systems and theory of superconductivity, including the high-$T_{c}$ case.}}
\author[V.M.~Loktev,
        V.~Turkowski]{V.M.~Loktev\refaddr{label1},
        V.~Turkowski\refaddr{label2}}
\addresses{
\addr{label1}Bogolyubov Institute for Theoretical Physics,
Metrologichna str. 14-b, Kyiv, 03680 Ukraine,\\
Electronic address: vloktev@bitp.kiev.ua \addr{label2}Department of
Physics and Astronomy, University of Missouri, Columbia, Missouri
65211, The USA,\\
Electronic address: turkowskiv@missouri.edu}
\begin{document}

\maketitle

\begin{abstract}
We analyze the spectral properties of a phenomenological model for a
weakly doped two-dimensional antiferromagnet, in which the carriers
move within one of the two sublattices where they were introduced. Such
a constraint results in the free carrier spectra with the maxima at
${\bf k}=(\pm \pi /2 ,\pm \pi /2)$ observed in some cuprates. We consider the spectral
properties of the model by taking into account fluctuations of the
spins in the antiferromagnetic background. We show that such fluctuations
lead to a non-pole-like structure of the single-hole Green's
function and these fluctuations can be responsible for some anomalous
"strange metal" properties of underdoped cuprates in the nonsuperconducting
regime.
\pacs 74.20.-z,74.20.Fg, 74.20.Rp, 74.72.-h
\end{abstract}

\section{Introduction}
\label{Introduction}

After more than twenty years from its discovery, the problem of high-temperature
superconductivity (HTSC) remains unsolved. Nevertheless, there are some
facts about HTSCs, which are generally accepted by the scientific community.
In particular, it is well known that these materials
transform from antiferromagnetic insulators into superconductors
with carrier doping. The superconductivity in most of the cuprates
mainly takes place in the ${\rm
CuO_{2}}$ layers, and the other inter-layer atoms supply the carriers for these
layers and play a role of the carrier scatterers. It is believed that the presence
of the antiferromagnetic background strongly affects the physical behavior
of the weakly doped materials. In particular, this behavior can be defined
by strong hole or electron correlations \cite{Franck,Shen}. Moreover, the correlations
can be responsible for the $d$-wave superconductivity in many cuprates
\cite{Ding,Loeser}. In fact, since the isotope effect in optimally doped HTSCs is rather weak,
it suggests that the electron-phonon coupling is not the main source of
superconductivity in these materials, though the role of phonons and, in particular,
 the interplay of strong correlations and
 phonon coupling in HTSCs is currently an active area of research (see, for
 example, Ref.~\cite{Macridin} and references therein).
Thus, it is believed by a significant part of the researchers that the phenomenon of HTSC can be
explained by using a strongly correlated model, in which the superconducting pairing
with the anisotropic order parameter is caused by an antiferromagnetic spin-wave coupling.
Probably, the most popular models for this scenario are the
two-dimensional Hubbard model and its approximation in the case of strong correlations, the $tJ$-model
\cite{Anderson,Zhang} (see, e.g.,
Refs.~\cite{Kane,Martinez,Plakida,Bensimon,Haule}).

Unfortunately, these models cannot be solved exactly in the two-dimensional case,
so it is difficult to make a firm conclusion whether they can be considered
as realistic models of HTSCs. As an alternative approach, one can consider a
simplified phenomenological model of cuprates with strong
correlations, which takes into account their main
properties and can be solved exactly. Typical unusual and important properties of
cuprates include a universal dome shape of the critical temperature - doping
curve, a different from the BCS theory ratio $2\Delta (T=0)/T_{c}^{max}\simeq 5.5$
at optimal doping, different critical densities
for the superconducting gap in the nodal and antinodal directions.
Probably, the most unusual phenomenon in HTSC is the
pseudogap phase in the underdoped regime. In this phase, the materials demonstrate
very unusual properties of a "strange metal", like an anomalous temperature dependence of the resistivity etc,
which are different from a Fermi-liquid behavior.
There is no general agreement in the HTSCs community on the origin of this phenomena,
however a recent improvement of the angle-resolved photoemission spectroscopy(ARPES) technique
gives a hope that the mystery of the "strange metal" phase in underdoped cuprates will be
resolved soon. Another unusual feature of some cuprates is their free carrier spectrum.
For example, the free spectrum of
 ${\rm Sr_{2}CuO_{2}Cl_{2}}$ and some other materials in
the insulating phase has the maxima at the momenta ${\bf k}=(\pm \pi /2 ,\pm \pi
/2)$ \cite{Damascelli}. This fact suggests that the main hopping
processes in these systems take place between the next nearest neighbor (NNN) and next NNN sites.
In other words, they correspond to an inter-sublattice carrier motion.
Since the NNN and next NNN hopping
parameters are too small for the oxygen sublattice of the ${\rm
CuO_{2}}$ planes, it is difficult to believe that
the holes move within the oxygen sites, as it is assumed by many
researchers \cite{Emery}. It is also difficult to assume that the free carrier
spectrum corresponds to the copper site NNN and next NNN hoping of the Zhang-Rice singlet,
 formed by
an oxygen hole and by one of the copper ions.
In fact, the Zhang-Rice singlet states, which move within their magnetic sublattice, are
unstable due to the hole frustration with respect
to the choice of the axis of the spin quantization and some other reasons
 \cite{Loktev,two_sublattice_phasediagram}.
In
order to describe some of the physical properties of the cuprates with the maximum of the free carrier
spectrum at ${\bf k}=(\pm \pi /2 ,\pm \pi
/2)$, there was proposed a model, in which the holes occupy and move within the copper
ion antiferromagnetic sublattice where they were born.\cite{Loktev}
It was assumed that the hopping takes place between the NNN and next NNN sites,
and the superconducting hole-hole attraction is due
to the minimization of the energy of the system, when two
holes occupy nearest sites. In this case, the minimal number of
the antiferromagnetic bonds is broken
\cite{Trugman,Wrobel1,Wrobel2,Wrobel3,Scalapino} and
similarly to the Hubbard and the ${\rm tJ}$-model cases,
the pairing takes places predominantly in the $d$-wave channel \cite{Scalapino}.

\begin{figure}[htb]
\centerline{\includegraphics[width=0.65\textwidth]{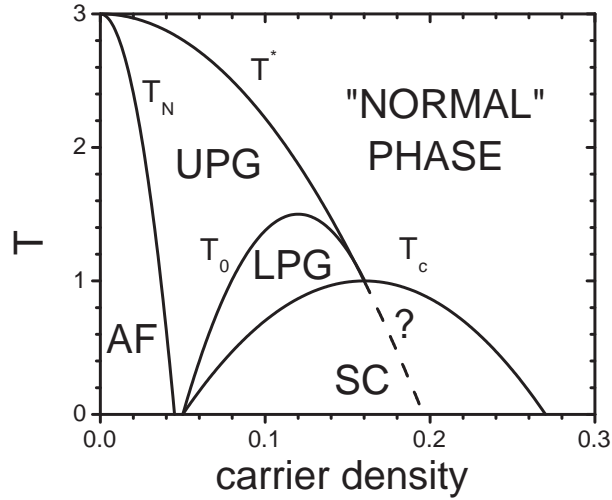}}
\caption{A typical temperature-carrier density phase diagram of the
hole-doped cuprates. It consists of the "normal", superconducting
(SC), antiferromagnetic (AF), and two pseudogap phase regions: the
lower pseudogap (LPG) phase, in which the behavior is affected by
both SC and AF spin phase fluctuations, and the upper pseudogap
(UPG) phase, in which only the AF spin phase fluctuations are
present.} \label{Fig1}
\end{figure}

We have already studied some of the properties of cuprates
in the superconducting and pseudogap phases in the framework of this model.
Namely, in Ref.~\cite{two_sublattice_phasediagram}, we have analyzed the
temperature-hole phase diagram of the model in the case of low
doping. In this paper, we have obtained the doping dependence
of the superconducting critical
temperature $T_{c}$ by solving a system of coupled equations for
the Green's function for the Hubbard operators
within a generalized mean-field approximation \cite{Mancini}. It was
shown that superconductivity in the model arises at finite doping and $T_{c}$
grows with doping in the underdoped regime. We have also shown that
there is an additional pseudogap phase at temperatures above $T_{c}$
and below another critical temperature $T_{0}$, which also grows with doping increasing
in the case of low carrier density (for a schematic picture, see Fig.~\ref{Fig1}).
Namely, according to the Emery-Kivelson
scenario \cite{Emery_Kivelson}, the Cooper pairs start to form below the temperature
$T_{0}$, which is associated with the
superconducting mean-field critical temperature $T_{c}^{MF}$.
In the two-dimensional case the pairs are disordered (the order parameter phases are exponentially
ordered) above the condensation temperature $T_{c}$.
This temperature can be associated with the
Berezinskii-Kosterlitz-Thouless (BKT) temperature, below which the
phases of the superconducting order parameter are algebraically
ordered. This is the only possible critical temperature in the two-dimensional case.
The existence of the superconducting pseudogap phase with unusual properties at
$T_{c}<T<T_{0}$ was confirmed in some cuprates, where a strong Nernst
effect was observed (see, for example Refs.~\cite{Lee,Wang},
and a theoretical paper \cite{Anderson2}). However, it is well-known that the anomalous properties
of cuprates in the underdoped regime take place up to the temperatures
much higher than $T_{0}$. In particular, an anomalous pseudogap in the one-hole
density of states is observed below a temperature $T^{*}\gg T_{0}$, which
is called the pseudogap critical temperature. We present a schematic phase
diagram of cuprates in Fig.~\ref{Fig1}, where we distinguish two regions in the pseudogap
phase: the superconducting lower pseudogap (LPG) phase and the other, which we call
the upper pseudogap (UPG) phase. The UPG critical temperature $T^{*}$ is a
decreasing function of doping. There are some experimental
evidences that $T^{*}$ goes below $T_{c}$ in the overdoped regime and approaches
zero at doping $\delta\simeq 0.19$.
It is believed by many researchers that the physical properties in
the region $T_{0}<T<T^{*}$ are defined by nonsuperconducting
processes (for over-view, see Ref.~\cite{Lee}). One of the popular
explanations is based on the idea of the spin singlet formations in
a doped two-dimensional antiferromagnet, which corresponds to the
resonant valence bond model. Unfortunately, there are no crucial
experimental results which confirm the existence of such a state so
far. Also, it is known that a non-Fermi liquid temperature dependence
of the conductivity in the underdoped cuprates
takes place up to temperatures of order 3000K, therefore, as it was suggested
by Phil Anderson \cite{Anderson1}, it shows that probably nonsuperconducting
effects are responsible for this phase. Namely, he suggested that the
unusual behavior of the spectral function can be explained by a renormalization
of the quasiparticle Green's function due to
a Gutzwiller projection in a strongly correlated model,
which leads to an additional time-dependence of the Green's function
and a non-pole-like (cut-like) form of this function in the frequency representation,
which corresponds to a non-Fermi-liquid case.

In paper \cite{Phys_Rev_B_submitted}, we have studied some of the spectral properties of the model
proposed in Ref.~\cite{Loktev} by taking into account fluctuations of the phases of the superconducting
order parameter and the phases of the spins in the antiferromagnetic background. In particular,
we have shown that the growth of the Fermi arcs with temperature in underdoped cuprates
can be qualitatively explained within the model by taking into account
fluctuations of the superconducting $d$-wave order parameter.
In this paper, we analyze the effect of the superconducting and spin fluctuations on the
structure of one-hole Green's function and the consequent anomalous behavior of the cuprates
in the pseudogap phase.
In particular, we show that a cut-like structure of the Green's function , qualitatively similar
to the one obtained
in Ref.~\cite{Anderson1}, can be obtained by taking into account these fluctuations.

The paper is organized as follows. The model and the main equations
in the mean-field case are presented in
Section \ref{Model}. In Section \ref{Spins}, we extend the problem
by taking into account fluctuations of the spins on the
antiferromagnetic sublattices and estimate the doping dependence of
the UPG critical temperature $T^{*}$. The results for
the spectral function and the density of states at different values
of temperature are presented in Sections \ref{Spectral} and
\ref{DOS}, correspondingly. In addition, in Section \ref{DOS} we show how
the anomalous frequency dependence of the conductivity can be obtained from the cut-like Green's
function. A summary, a discussion of the results
and conclusions are given in Section \ref{Conclusions}.

\section{Model}
\label{Model}

As it was shown in Ref.~\cite{Loktev}, the effective model for the
holes in some of the weakly doped cuprates can be written as:
\begin{eqnarray}
H=(\varepsilon_{d}-\mu)\sum_{{\bf n}}X_{{\bf n}}^{2,2}
-\frac{1}{2}\sum_{{\bf n},{\bf m}}t_{{\bf n}{\bf m}} \cos\frac{{\bf
Q}_{AFM}({\bf n}-{\bf m})}{2} X_{{\bf n}}^{2,1/2}X_{{\bf m}}^{1/2,2}
-J\sum_{{\bf n},{\bf \rho}={\bf a},{\bf b}} X_{{\bf n}}^{2,2}X_{{\bf
n}+{\bf \rho}}^{2,2}. \label{Hamiltonian2}
\end{eqnarray}
In the last equation, $X_{{\bf n}}^{2,2}$, $X_{{\bf n}}^{1/2,2}$ $X_{{\bf
n}}^{2,1/2}$ are the Hubbard operators for the hole number, the hole
annihilation and the hole creation on the site ${\bf n}$. The
first three terms in the Hamiltonian Eq.~(\ref{Hamiltonian2}) describe
the local hole energy and the NN, and NNN hopping
processes, where $\varepsilon_{d}$ and $\mu$ are the hole
on-site energy and the chemical potential. The magnetic structure
vectors ${\bf Q}_{AFM}$ are equal to $(\pm \pi ,\pm \pi )$, which
corresponds to the antiferromagnetic case.
We use the local spin coordinates for the Hubbard operators.
The noninteracting part of the hole Hamiltonian describes
holes, which move within their sublattices. The free hole dispersion
relation is
\begin{equation}
\varepsilon ({\bf k})=\varepsilon_{d} -4t_{2}\cos k_{x}\cos k_{y}
-2t_{3}(\cos 2k_{x}+\cos 2k_{y})-\mu . \label{epsilon}
\end{equation}

The effective hole-hole attraction in the system is described by the last term
in Eq.~(\ref{Hamiltonian2}). In fact, the doped holes introduced on the
antiferromagnetic lattice, which move within the sublattice they were introduced,
lead to a minimal increasing of the energy of the system
when they sit on the nearest sites \cite{Trugman}, since in this case the minimal number
of the antiferromagnetic couplings $J$ between the nearest site spins is
broken. In this case, two doped holes will always try to occupy NN sites,
which results in the effective attraction described by the last term of the Hamiltonian
Eq.~(\ref{Hamiltonian2}). In our calculations, we use the length units such that
the lattice constant is equal to one, $a=1$, and choose the energy parameter
$\varepsilon_{d}$ to be equal to $4t_{2}+4t_{3}$.
In this case, the free hole energy is equal to zero at ${\bf k}=0$ in the limit of low doping
($\mu \rightarrow 0$).
In order to find dependence of the physical properties of the system
on hole concentration $\delta$, we shall use the following equation, which
defines $\delta$ in terms of the Hubbard particle number operator:
\begin{equation}
\delta =\sum_{{\bf n}}\langle X_{{\bf n}}^{2,2}\rangle .
\label{particlenumber}
\end{equation}

Recently, we have studied the spectral properties of the model
my taking into account superconducting fluctuations \cite{Phys_Rev_B_submitted}.
In this paper, we use a similar formalism to study the effect of the antiferromagnetic
background spin fluctuations on the anomalous spectral and some other properties
of the system in the pseudogap phase at $T>T_{c}$. Despite we are interested mainly
in the spin fluctuation effects, we shall consider the general case assuming that the superconducting
 pairing can also take place, which corresponds to the temperature interval
$T_{c}<T<T_{0}$. At higher temperatures, i.e. at $T_{0}<T<T^{*}$, we shall use the same
equations by putting the superconducting gap to be equal to zero in our calculations.

Then, in order to study the properties of the system described by the Hamiltonian Eq.~(\ref{Hamiltonian2}),
it is convenient to
introduce generalized Nambu-Hubbard hole operators
\begin{equation}
\Psi_{{\bf n}}(t)= \left(
\begin{array}{c}
X_{{\bf n}}^{2,1/2}(t)\\
X_{{\bf n}}^{1/2 ,2}(t)
\end{array}\right) , \ \ \ \
\Psi_{{\bf n}}^{\dagger}(t)= \left( X_{{\bf n}}^{1/2 ,2}(t), X_{{\bf
n}}^{2,1/2}(t) \right) , \label{Nambu}
\end{equation}
where ${\bf n}$ is the lattice site and $t$ is time,
and to calculate the time-ordered Green's function
\begin{eqnarray}
{\hat G}_{{\bf n}{\bf m}}(t,t')= -i\langle T (\Psi_{{\bf
n}}(t)\Psi_{{\bf m}}^{\dagger}(t'))\rangle . \label{GFdefinition}
\end{eqnarray}
This function satisfies the following equation:
\begin{equation}
i\frac{\partial}{\partial t} {\hat G}_{{\bf n}{\bf m}}(t,t') =\delta
(t-t')\delta_{{\bf n}{\bf m}}{\hat I} +\langle T[\Psi_{{\bf
n}}(t),H]\Psi_{{\bf m}}^{\dagger}(t') \rangle , \label{G}
\end{equation}
where ${\hat I}$ is a diagonal $2\times 2$-matrix with the nonzero elements equal to
$\langle X_{{\bf n}}^{1/2,1/2}(t)+X_{{\bf n}}^{2,2}(t)\rangle$. In the case of low doping,
${\hat I}\simeq {\hat 1}$.
In order to solve Eq.~(\ref{G}), it is convenient to approximate its last term by a generalized
mean-field theory expression:
\begin{equation}
\langle T[\Psi_{{\bf n}},H]\Psi_{{\bf m}}^{\dagger}\rangle (\omega
)\simeq \sum_{{\bf l}}{\hat E}_{{\bf n}{\bf l}}{\hat G}_{{\bf l}{\bf
m}}(\omega ), \label{linearization}
\end{equation}
where
\begin{equation}
{\hat E}_{{\bf n}{\bf m}}=\langle \{ [\Psi_{{\bf n}},H], \Psi_{{\bf
m}}^{\dagger}\}\rangle  \label{linearization2}
\end{equation}
is the energy matrix (for details see, e.g., Ref.~\cite{Mancini}).
In this approximation, one neglects the
dynamical corrections to the self-energy, which can be systematically
taken into account. For example, in Ref.~\cite{Plakida} the authors
considered these corrections in the case of the ${\rm
tJ}$-model by using a similar formalism.
In order to find the Green's function, one needs to calculate the elements
of the energy matrix ${\hat E}_{{\bf n}{\bf l}}$, which depend on different
correlation functions, in particular on the superconducting gap function,
which can be found by using the fluctuation-dissipation theorem.
Assuming that the superconducting pairing takes place
in the d-wave channel and introducing the relevant gap function:
\begin{eqnarray}
\Delta_{d} ({\bf k})= -4J\sum_{{\bf q}} \gamma_{d}({\bf
k})\gamma_{d}({\bf q}) \langle X_{-{\bf q}}^{2,1/2}X_{{\bf
q}}^{2,1/2} \rangle \equiv \Delta_{d} \gamma_{d}({\bf k}),
\label{Deltak}
\end{eqnarray}
where $\gamma_{d}({\bf k})=\cos (k_{x})-\cos (k_{y})$ is a $d$-wave
structure factor, one can get the following approximate expression for
the Green's function in the frequency-momentum representation:
\begin{equation}
G(\omega ,{\bf k})=\frac{1}{\omega +\varepsilon ({\bf k})\tau_{z}
+i\Delta({\bf k})\tau_{y}}, \label{GF}
\end{equation}
where ${\hat \tau}_{y}$ and ${\hat \tau}_{z}$ are the Pauli
matrices.

In order to find the doping and temperature dependencies of the
superconducting gap parameter $\Delta_{d}$, one needs to derive
and to solve the system of equations for $\Delta_{d}$ and the chemical
potential $\mu$. These equations follow from the definitions
Eqs.~(\ref{particlenumber}), (\ref{Deltak}) and the self-consistency
conditions, which follow from the fluctuation-dissipation theorem:
\begin{equation}
1=4J\sum_{\bf q}\gamma_{d}^{2}({\bf q}) \tanh \left(
\frac{\sqrt{\varepsilon^{2}({\bf
q})+\Delta_{d}^{2}\gamma_{d}^{2}({\bf q})}}{2T}\right) \frac{1}
{\sqrt{\varepsilon^{2}({\bf q})+\Delta_{d}^{2}\gamma_{d}^{2}({\bf
q})}}, \label{gapequationT}
\end{equation}
\begin{equation}
\delta =\sum_{{\bf q}}\left[1+\tanh \left(
\frac{\sqrt{\varepsilon^{2}({\bf
q})+\Delta_{d}^{2}\gamma_{d}^{2}({\bf
q})}}{2T}\right)\frac{\varepsilon ({\bf q})}
{\sqrt{\varepsilon^{2}({\bf q})+\Delta_{d}^{2}\gamma_{d}^{2}({\bf
q})}}\right] . \label{numberequationT}
\end{equation}
 (see Ref.~\cite{two_sublattice_phasediagram} for details). The solution of these
 equations at $\Delta_{d}=0$ gives the doping dependence of the
 mean-field critical temperature $T_{c}^{MF}$, or the LPG
 critical temperature $T_{0}$. According to the Emery-Kivelson scenario, the real superconducting
critical temperature $T_{c}<T_{c}$ in the two-dimensional case corresponds to the BKT
 temperature, below which the phases of the order parameter become
 algebraically ordered (for over-review see,
 for example, Ref.~\cite{PhysRep}). We studied the doping dependence of this temperature
in Ref.~\cite{two_sublattice_phasediagram}. In this paper, we shall mainly concentrate
on the spectral properties of the model in the pseudogap phase at $T>T_{c}$
by taking into account superconducting fluctuations and spin fluctuations of the antiferromagnetic
background.

\section{Spin fluctuations}
\label{Spins}

In order to describe the UPG region, we assume that, in
analogy with the superconducting order parameter fluctuations, the
physics in the temperature range $T_{0}<T<T^{*}$ is governed by the
fluctuations of the spin phases. Supposing that the antiferromagnetic
copper oxide spin Hamiltonian is described by the XY-model, it is
easy to obtain an analogous critical temperature for the spin
subsystem. Therefore, we associate $T^{*}$ with the temperature of
the BKT transition for spins. This temperature can be also estimated
from the following equation:
\begin{equation}
T^{*}=\frac{\pi}{2}{\cal J}_{spin}(\mu , T^{*}), \label{TBKTspin}
\end{equation}
where the spin stiffness ${\cal J}_{spin}$ is the coefficient
in front of the quadratic term of the spin phase gradients
in the effective action for the spin phase $\varphi_{{\bf n}}={\bf Q}_{AFM}{\bf n}$
differences:
\begin{equation}
\Omega = \frac{{\cal K}_{spin}}{2}\int d^{2}r \varphi
\partial_{t}^{2}\varphi +\frac{{\cal J}_{spin}}{2}\int d^{2}r
({\bf\nabla}\varphi )^{2}. \label{Omegaspin}
\end{equation}
It is equal to
\begin{eqnarray}
{\cal J}_{spin}=\frac{1}{2}\int \frac{d^{2}k}{(2\pi )^{2}} \tanh
\left( \frac{\sqrt{{\bar \varepsilon}^{2}({\bf
k})+\Delta_{d}^{2}({\bf k})}}{2T}\right) \frac{{\bar
\varepsilon}^{2} ({\bf k})}{\sqrt{{\bar \varepsilon} ({\bf
k})^{2}+\Delta_{d}^{2}({\bf k})}} , \label{Jspin}
\end{eqnarray}
where ${\bar \varepsilon}({\bf k})=\varepsilon ({\bf k})-4J\delta
+4t_{1}\delta (\cos k_{x}+\cos k_{y})$ (see
\cite{Phys_Rev_B_submitted}). As it was mentioned above, doping leads to a
gradual destruction of the antiferromagnetic order and to a hole
hopping between NN sites, which belong to different sublattices. We
take into account this process by adding the doping-dependent term
$4t_{1}\delta (\cos k_{x}+\cos k_{y})$ to the free spectrum
Eq.~(\ref{epsilon}). The doping dependence of $T^{*}$ can be found
from Eq.~(\ref{TBKTspin}) at $\Delta_{d}=0$. In this case,  Eq.~(\ref{TBKTspin})
 has the following form:
\begin{eqnarray}
T^{*}=\frac{\pi}{4}\int \frac{d^{2}k}{(2\pi )^{2}} \tanh \left(
\frac{ {\bar \varepsilon}({\bf k})}{2T^{*}}\right) {\bar
\varepsilon} ({\bf k}). \label{TBKTspin2}
\end{eqnarray}
The exact solution of Eq.~(\ref{TBKTspin2}) shows that $T^{*}$ is a
decreasing function of doping. One can estimate the analytical
dependence $T^{*}(\delta)$ from Eq.~(\ref{TBKTspin2}) by taking into
account that at low doping $T^{*}\gg |{\bar \varepsilon}({\bf k})|$.
Therefore, in this case:
\begin{eqnarray}
T^{*}\simeq\sqrt{\frac{\pi}{8}\int \frac{d^{2}k}{(2\pi )^{2}} {\bar
\varepsilon}^{2} ({\bf k})}\simeq T_{max}^{*}(1-\beta\delta ),
\label{TBKTspin3}
\end{eqnarray}
where $T_{max}^{*}=\sqrt{\frac{\pi}{8}\int \frac{d^{2}k}{(2\pi
)^{2}} {\bar \varepsilon}^{2} ({\bf k})}|_{\delta =0}$ and $\beta
=4\sqrt{\pi /2}t_{1}\int\frac{d^{2}k}{(2\pi )^{2}}\varepsilon ({\bf
k})(\cos k_{x}+\cos k_{y})$.

\section{Spectral function}
\label{Spectral}

In order to study the spectral function in the case when the spin phase fluctuations
are taken into account, it is convenient to write down the spin
functions explicitly in the laboratory system of coordinates:
\begin{eqnarray}
|\sigma_{{\bf n}}\rangle=|\sigma\rangle \cos (\varphi_{{\bf n}}/2 )
+2\sigma |\sigma\rangle \sin (\varphi_{{\bf n}}/2 ) ,
\label{spinphases}
\end{eqnarray}
where $\varphi_{{\bf n}}$ is the angle between the directions of the
laboratory and local systems of coordinates. For simplicity, we
assume that the spins lie in the plane of the system and their phase fluctuations are small.
In this case, one can get the following result for the Green's function:
$$
G_{\alpha \beta}(x)={\cal
G}_{\alpha \beta}(x){\cal D}^{spin}(x),
$$
where ${\cal G}_{\alpha\beta}(x)$ is the Green's function of the
Nambu spinors
and
$$
{\cal D}^{spin}(x) =
e^{-\langle \varphi (x)\varphi (0)\rangle /4}
$$
is the correlator of the spin phases (for details, see Refs.~\cite{Phys_Rev_B_submitted,PhysRep,Patashinskii}).
The phase correlators in the last two expressions can be easily obtained from the
effective action for the spin phases. Namely,
\begin{eqnarray}
D^{spin}(\tau ,{\bf r})=\exp\left[ -\frac{T}{4}
\sum_{n=-\infty}^{\infty} \int\frac{qdqd\varphi}{(2\pi )^{2}}
\frac{1-\cos (qr\cos \varphi )\cos (\Omega_{n}\tau )}{{\cal
J}_{spin}q^{2}+{\cal K}_{spin}\Omega_{n}^{2}} \right] ,
\label{Dspindynamic}
\end{eqnarray}
where ${\cal J}_{spin}$ and ${\cal
K}_{spin}$ are the coefficients in front of the gradient and time
derivative terms in the effective action Eq.~(\ref{Omegaspin}).
One can use these exact expressions to calculate the spin correlation
functions. However, in order to study the qualitative behavior of the spectral function
of the system, it is enough to approximate the correlation function by
\begin{eqnarray}
D_{spin}(x)=\left[\theta (T^{*}-T) +\theta (T-T^{*})\exp
(-r/\xi_{spin}(T))\right]
e^{-\Gamma_{spin} t}(r/r_{spin})^{-\alpha_{spin} }, \label{Dspin}
\end{eqnarray}
where $T^{*}$ is the spin BKT temperature and $\alpha_{spin}$, $\xi_{spin}$,
$r_{spin}$ and $\Gamma$ are doping- and temperature-dependent
parameters. In principle, one also needs to include superconducting fluctuations
in order to describe the properties of the system at $T_{c}<T<T_{0}$.
In this case, $D^{spin}$ must be multiplied by the corresponding superconducting
function $D^{SC}$ with the space-time dependence similar to Eq.~(\ref{Dspin}).
We assume that $D^{spin}$ describes the total effect of the spin and superconducting
fluctuations and use the corresponding parameter notations $\alpha , \xi ,
r$ and $\Gamma$ instead of ones used in Eq.~(\ref{Dspin}).
The values of the parameters can be estimated from experiments (see below).

In the frequency-momentum representation, the Green's function
has the following form:
\begin{eqnarray}
G(i\omega_{n},{\bf k})=-T\sum_{m}\int d {\bf
q}\frac{i\omega_{m}+\tau_{3}\varepsilon ({\bf q})} {\omega_{n}^{2}
+\varepsilon^{2}({\bf q})+\Delta_{d}^{2}({\bf q})} {\cal D}^{spin} ({\bf
k}-{\bf q},\omega_{n}-\omega_{m}), \label{GMatsubara}
\end{eqnarray}
where the Fourier transform of the correlator of the phase
fluctuations ${\cal D}^{spin}({\bf k}-{\bf q},\omega_{n}-\omega_{m})$ can
be found from Eq.~(\ref{Dspin}). In particular, at
$T<T^{*}$:
\begin{eqnarray}
D^{spin}(i\Omega_{n},{\bf q})=A\frac{1}{[{\bf
q}^{2}+\xi^{-2}(T)]^{\alpha}}
\frac{\Gamma}{\omega_{n}^{2}+\Gamma^{2}} , \label{DSCMatsubara}
\end{eqnarray}
where $A$ is a parameter.
We shall use this expression to analyze the spectral properties of
the system. For simplicity, we assume that the
inverse time correlation length $\Gamma$ is proportional to temperature
and put $\Gamma =0.1T/{\bar T}$, where ${\bar T}\sim T^{*}$.

The spectral properties of the holes can be studied by making the
analytical continuation $i\omega_{n}\rightarrow \omega +i\eta$
($\eta\rightarrow +0$) and extracting the imaginary part of the
Green's function
\begin{eqnarray}
A(\omega ,{\bf k})=-(i/\pi){\rm Im}G(\omega ,{\bf k}). \label{A}
\end{eqnarray}

In this case, by using Eqs.~(\ref{GMatsubara})-(\ref{A}) and performing
one frequency integration, one can get
the following expression for the spectral function:
\begin{eqnarray}
A(\omega ,{\bf k})=A\int d{\bf q} \left\{ \left[ 1+\frac{\xi ({\bf
q})}{E({\bf q})} \right] \frac{\Gamma}{(\omega -E({\bf
q}))^{2}+\Gamma^{2}} +\left[ 1-\frac{\xi ({\bf q})}{E({\bf q})}
\right] \frac{\Gamma}{(\omega +E({\bf
q}))^{2}+\Gamma^{2}}\right\} \nonumber \\
\times \frac{1}{\left[ ({\bf k}-{\bf
q})^{2}+\xi^{-2}(T)\right]^{\alpha}},
 \label{A3}
\end{eqnarray}
where $A$ can be most easily found by using the Green's function zeroth spectral moment sum rule.
Let us get an approximate expression for the spectral function
in the case of $\Delta_{d}=0$ and positive frequencies. In this case, the last term in the
figure brackets can be neglected, and one gets
\begin{eqnarray}
A(\omega ,{\bf k})\simeq {\bar A}(\omega)\int d\varphi_{q}
\frac{1}{\left[ ({\bf k}^{2}-2k\sqrt{2m^{*}(\mu +\omega )}\cos (\varphi_{q})
+ 2m^{*}(\mu+\omega )+\xi^{-2}(T)
\right]^{\alpha}},
 \label{A4}
\end{eqnarray}
where
$$
{\bar A}(\omega )=Am^{*}\int_{0}^{\infty} d\xi [ 1+ {\rm sign} (\xi
-\mu )] \frac{\Gamma}{(\omega -\xi )^{2}+\Gamma^{2}}\simeq 2
m^{*}A\left[ \frac{\pi}{2} +\arctan (\omega /\Gamma ) \right].
$$
As it follows from Eq.~(\ref{A4}), the Green's function has a
cut-like form, contrary to the Fermi-liquid pole-like case. Such a
dependence is caused by presence of the spin correlation function
${\cal D}^{spin} ({\bf k}-{\bf q},\omega_{n}-\omega_{m})$ in
Eq.~(\ref{GMatsubara}), which smoothes out the $\delta$-function
peaks of the spectral function that come from the denominator of the
fermion Green's function. Therefore, the quasi-particle residue $Z$,
which can be defined as the coefficient in front of the Fermi
quasi-particle spectral function $\delta$-peak, is equal to zero.
This means that the system is in a non-Fermi-liquid regime. This
result is qualitatively similar to the result recently obtained by
P.W.~Anderson in the case of a strongly correlated model
\cite{Anderson1} and used to describe the spectral function in
$Bi_{2}Sr_{2}CaCu_{2}O_{8+\delta}$ in Ref.~\cite{Anderson3}:
\begin{eqnarray}
A(\omega ,{\bf k})=f\left(\frac{\omega}{T}\right)
\frac{\sin [(1-p)(\pi /2-\tan^{-1}[(\omega -v_{F}k)/\Gamma])]}{[(\omega -v_{F}k)^{2}+\Gamma^{2}]^{(1-p)/2}},
\label{A_Anderson}
\end{eqnarray}
where $f(\omega /T)$ is the Fermi function and p and $\Gamma$ are parameters
in the effective Green's function:
\begin{eqnarray}
G(t,{\bf x})=
\frac{t^{-p}e^{-\Gamma t}}{|{\bf x}|-v_{F}t}.
\label{G_Anderson}
\end{eqnarray}
In particular, the pre-factor $t^{-p}$ comes from the contribution of the Gutzwiller projection
on the single occupied states of the strongly correlated system. It was estimated that $p\simeq 0.12$
and $\Gamma =AT+B(k-k_{F})^{2}$ in the momentum space, where $A$ and $B$ are constants.

In fact, in our case we can get an approximate cut-like expression for the spectral function
by putting $\varphi_{q}=0$ in the expression under the integral in Eq.~(\ref{A4})
and integrating over the angle $\varphi_{q}$. This can be done since the momentum angle region around
$\varphi_{q}=0$ gives the largest contribution into the integral. In this case,
\begin{eqnarray}
A(\omega ,{\bf k})\simeq 4\pi m^{*}A\left[ \frac{\pi}{2}+\arctan
(\omega /\Gamma ) \right] \frac{1}{\left[ (k-\sqrt{2m^{*}(\mu
+\omega)})^{2}+\xi^{-2}(T) \right]^{\alpha}}.
 \label{A5}
\end{eqnarray}

\begin{figure}[htb]
\centerline{\includegraphics[width=0.65\textwidth]{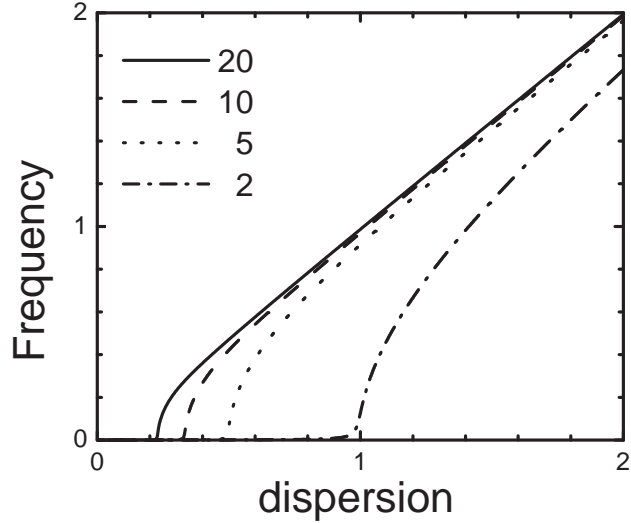}}
\caption{ The spectral function peak frequency $\omega$-dispersion
$\xi ({\bf k})$ curve at different values of $\alpha$. This curve
approaches the free fermion curve $\omega =\xi ({\bf k})$ as
$\alpha$ increases.} \label{Fig2}
\end{figure}

As it follows from this equation, the spectral function has the maximum at $\omega =\xi ({\bf k})
=k^{2}/(2m^{*})$, similar to the free hole case (we put $\mu =0$ for simplicity), but
this function is a smooth function, different from the delta-function. Our numerical evaluation
of the integral in Eq.~(\ref{A4}) show that the approximation Eq.~(\ref{A5}) is correct
only at large $\alpha$ (Fig.~\ref{Fig2}).
In the case of small $\alpha$, the $\omega [\xi ({\bf k})]$-dependence for the spectral function
maximum is linear at large $\xi ({\bf k})$. In this case, the curve begins at finite value of $|{\bf k}|$,
which decreases with $\alpha$ increasing.

From Eq.~(\ref{A5}), one can find an approximate expression for the
spectral function at $\omega =0$:
\begin{eqnarray}
A(\omega =0,{\bf k})\simeq
2\pi^{2} m^{*}A\frac{1}{\left[ (k-k_{F})^{2}+\xi^{-2}(T)
\right]^{\alpha}}.
 \label{A6}
\end{eqnarray}
As it follows from this equation, the spectral weight on the Fermi level is defined
by $\xi$ and $\alpha$.

\section{Density of states}
\label{DOS}

The density of states can be obtained from Eq.~(\ref{A3}) by using
the standard expression
\begin{eqnarray}
N(\omega)=\int \frac{d^{2}k}{(2\pi )^{2}}A(\omega ,{\bf k}).
\label{N}
\end{eqnarray}
Similar to the previous Section, one can show that the system
demonstrates a finite DOS at zero frequency and it is defined by the
parameters $\xi$ and $\alpha$.

One can estimate the DOS weight at $\omega =0$ from Eq.~(\ref{A6}):
\begin{eqnarray}
N(\omega =0)
&\simeq&
2\pi^{2} m^{*}A \int \frac{d^{2}k}{(2\pi )^{2}}
\frac{1}{\left[ k^{2}+\xi^{-2}(T)
\right]^{\alpha}}
\nonumber \\
&\simeq& \frac{\pi m^{*}A}{2(1-\alpha)\xi^{2(1-\alpha )}}
\left[(2m^{*}W\xi^{2}(T)+1)^{1-\alpha}-1\right] .
 \label{N2}
\end{eqnarray}
From Eqs.~(\ref{A5}), (\ref{A6}) and (\ref{N2}) one can estimate the values of the phenomenological
parameters $\xi , \alpha$ and $\Gamma$ by comparing theoretical results with experimental data. In particular,
one can get that $\Gamma \sim T$ and $\xi$ is a weakly-dependent function of temperature
for a wide temperature range above $T_{0}$.

\begin{figure}[htb]
\centerline{\includegraphics[width=0.65\textwidth]{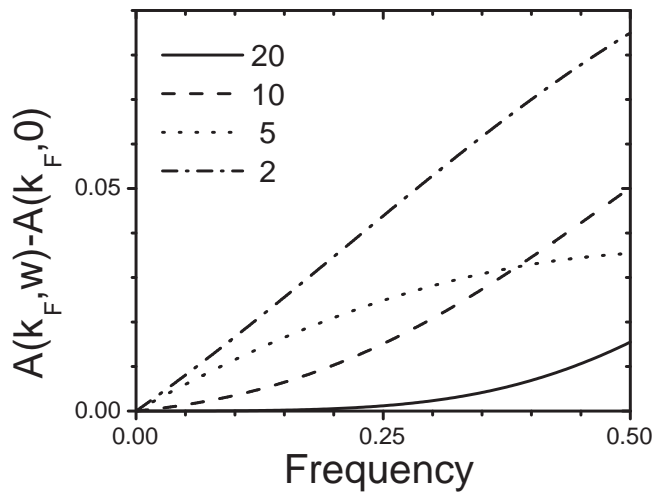}}
\caption{ The leading frequency-dependent term of the spectral
function (in arbitrary units) at small $\omega$, $k=k_{F}$ and
different values of $\alpha$. This term is proportional to
$A(k_{F},\omega) -A(k_{F},0)$. The frequency $\omega$ is given in
units of $1/(2m^{*}\xi^{2})$. } \label{Fig3}
\end{figure}

To conclude this Section, we would like to demonstrate how some of the non-Fermi-liquid properties
could result from the cut-like structure of the Green's function by using the anomalous conductivity
as an example. One can roughly estimate the conductivity to be proportional to the quasiparticle
life-time $\tau ({\bf k},\omega )$ at $|{\bf k}|=k_{F}$. This quantity can be estimated
to be inversely
proportional to the imaginary part of the one-hole self-energy, i.e.
$\tau (k_{F},\omega )\sim 1/Im \Sigma (k_F,\omega)$. On the other hand, since
$$
A({\bf k},\omega) =\frac{Im\Sigma ({\bf k},\omega)}{ ({\bf k}^2/2m^{*}-\mu)^{2}
+Im\Sigma^{2} ({\bf k},\omega)},
$$
where we have neglected the real part of the self-energy, one immediately gets
$\tau (k_{F},\omega )\sim A(k_F,\omega)$.
The frequency dependence of this quantity at different values of $\alpha$
is presented in Fig.~\ref{Fig3}, where we have substracted the frequency-independent part from the spectral function.
As it follows from this Figure, at low and moderate frequencies the dependence of the conductivity
on frequency can be approximated by $\sigma \sim \omega^{b}$, where $b$ is a parameter.
As it follows from Eq.~(\ref{A4}), at small frequencies
\begin{eqnarray}
A(k_{F},\omega)-A(k_{F},\omega)\sim \frac{\alpha^{2}\xi (k_{F})
/\xi (T)-\alpha}{(\xi (k_{F})/\xi (T)+1)^{\alpha +2}}\omega .
\label{Asmallw}
\end{eqnarray}
Such a dependence obtained by using a rather rough approximation already indicates that the cut-like form
of the Green's function can result in a non-Fermi-liquid behavior of the system.

In order to make a comparison with experiments for conductivity and other quantities in the pseudogap phase,
one needs to take into account more accurately different properties of the materials, like the band structure,
the antiferromagnetic spin coupling, which defines $J$ and others. Such a comparison is also necessary
to the phenomenological parameters for the spin fluctuation correlation function Eq.~(\ref{Dspin}).

\section{Conclusions}
\label{Conclusions}

In this paper, we have considered the spectral properties of a
phenomenological model of HTSC in the underdoped regime by taking
into account fluctuations of the phases in the AF spin background.
Namely, we have considered the temperature
evolutions of the spectral function and of the density of states.
By studying the spectral function, we
have shown that its temperature dependence in the case of HTSCs can be qualitatively described
by this model in the case of the
proper choice of the decoherence time correlation length and other parameters for the spin angle
correlation function.
These parameters can be taken from experiments and they are directly connected
with the microscopic model parameters. Similarly, we have derived and analyzed the expression
for the density of states. Finally, we have shown that
the spin fluctuations can be responsible for the anomalous behavior of the conductivity
in the underdoped regime.
We have compared our result for the Green's function with the expression proposed in \cite{Anderson1},
 and have shown that
both models can describe anomalous properties of underdoped cuprates
without using exotic models, like the marginal Fermi-liquid etc. The validity of the model
studied in this paper to describe all properties of cuprates in the pseudogap phase can be tested
by taking into account different experimental phenomena, what requires a farther investigation.

\section*{Acknowledgments}

V.M.L.~would like to thank P.W.~Anderson for a beneficial
communication. V.T.~acknowledges support by the National Science
Foundation under grant number DMR-0553485.

%
%


%
%
%



\end{document}